\documentclass{article}
\usepackage{amsmath}
\usepackage{bigints}
\usepackage{appendix}
\usepackage[numbers]{natbib}
\usepackage{graphicx} 
\usepackage{authblk}
\usepackage{amssymb}

\title{On the secular evolution of the semi-major axis in canonical formalism}
\author[1,2]{Barnab{\'a}s Deme}
\affil[1]{Baja Astronomical Observatory of SZTE University, Szegedi út, Kt. 766,
Hungary, H-6500}
\affil[2]{ELTE E{\"o}tv{\"o}s Lor{\'a}nd University, Institute of Physics and Astronomy, Department of Astronomy, P{\'a}zm{\'a}ny P{\'e}ter stny. 1/A, H-1117, Budapest, Hungary}

\begin{document}

\maketitle

\begin{abstract}
    There are several astrophysical configurations where one is interested only in the long-term dynamical evolution. Although the first-order version of this approximation is usually sufficient in applications, second-order corrections may be relevant, too. Here we use the Hamiltonian formalism to show how such higher-order terms lead to the long-term evolution of the semi-major axis.  
\end{abstract}

\section{Introduction}
The timescale of astrodynamical phenomena usually spans several orders of magnitude. In particular, one usually finds an oscillatory behavior on orbital times, superposed on a much slower variation of the orbital parameters. The latter is called secular evolution. In the Hamiltonian framework of classical mechanics, it is obtained by averaging the Hamiltonian over the fast angles. More precisely, one performs the so-called von Zeipel canonical transformation in order to eliminate the oscillating perturbations that do not accumulate on longer timescales \cite{vonzeipel10,arnold78}. 

The hierarchical three-body problem offers an excellent site to study secular evolution. It hosts the celebrated Kozai-Lidov mechanism \cite{kozai62,lidov62,ito19}, which may be responsible for several interesting phenomena ranging from planetary to black hole dynamics \cite{naoz2016,shevchenko17,perets2025}. In the standard quadrupole approximation, a sufficiently inclined triple exhibits eccentricity and inclination oscillations. The common wisdom is that the semi-major axis remains secularly constant, in a sense that its dynamics consists of orbital-time oscillations only.

Interesting new dynamics may appear in a second-order theory, i.e. if one takes into account the back-reaction of the quadrupole perturbation on itself (see Ref.~\cite{tremiane2023} and references therein). For example, the argument of periapsis may be perturbed significantly compared to the argument of ascending node (see Sec. 8-3. in Ref.~\cite{hestenes02}), a fact which seriously challenged Newton's gravitational theory in the $18^\mathrm{th}$ century \cite{bodenmann2010}. Also, quadrupole-squared effects may quench flips from prograde to retrograde orbits \cite{conway2025}, which are characteristic for the first-order (octupole)\footnote{Note that first-order here refers to the fact that the given term's self-interaction is not taken into account.} Kozai-Lidov mechanism \cite{naoz2013}. Moreover, contrary to the standard viewpoint that holds in the first-order approximation mentioned above, the semi-major axis may no longer be constant secularly, i.e., it may oscillate on timescales much longer than the orbital ones. This surprising result was recently derived in Ref.~\cite{conway2024a} using Lagrange's planetary equations \cite{valtonen06}, without referring to the Hamiltonian formalism explicitly. Ref.~\cite{li2025} provided a detailed Hamiltonian description of these dynamical subtleties; however, they did not comment on the interesting secular evolution of the semi-major axis. In the spirit of Poisson's 1809 theorem on the stability of the solar system, Sec. 8.9 of Ref.~\cite{hagihara1972} discusses a similar problem but puts the focus on the absence of terms proportional to time. Sec. 5.1 of Ref.~\cite{tremaine2023book} hints at the presence of secular modulations of the semi-major axis (see its footnote 1), but calls it slow oscillation and uses it in a slightly different context.

The aim of this work is to show how the secular evolution of the semi-major axis can be derived in the Hamiltonian framework. This work is structured as follows. Sec.~\ref{sec:general} introduces the basic concepts. In Sec.~\ref{sec:toymodel} we apply these ideas to a simple toy model. In Sec.~\ref{sec:discussion} we conclude.

\section{General formalism}\label{sec:general}
Let us take a general Hamiltonian 
\begin{equation}
    \mathcal{H}=\mathcal{H}_0(L)+\epsilon \left(\overline{\mathcal{H}}_1(L,g,G)+\widetilde{\mathcal{H}}_1(l,L,g,G)\right),
\end{equation}
where we use Delaunay's variables and deal with the 2D case for the sake of simplicity. $\mathcal{H}_0$ is the integrable part, $\overline{\mathcal{H}}_1$ is the averaged, while $\widetilde{\mathcal{H}}_1$ is the oscillating perturbation ($\epsilon \ll 1$). In von Zeipel's canonical theory, if there are no resonances, the latter is eliminated using a generating function of the following general form \cite{harrington1968}:
\begin{equation}\label{eq:generator}
    S=\mathrm{id.}+\epsilon\sum_k S_k(g,L',G')\mathrm{e}^{\mathrm{i}kl},
\end{equation}
where $\mathrm{id.}$ is the identity transformation. The conversion between the old ($q,p$) and new canonical variables ($q',p'$) is given by \cite{goldstein02}
\begin{align}
    q'=\frac{\partial S}{\partial p'} && \mbox{and} && p=\frac{\partial S}{\partial q}.
\end{align}
Applying it to \eqref{eq:generator} and inverting them lead to
\begin{equation}\label{eq:l}
    l=l'+\epsilon\sum_k \frac{\partial S_k(g',L',G')}{\partial L'}\mathrm{e}^{\mathrm{i}kl'}+\mathcal{O}(\epsilon^2),
\end{equation}
\begin{equation}\label{eq:L}
    L=L'+\epsilon\sum_k \mathrm{i}kS_k(g',L',G')\mathrm{e}^{\mathrm{i}kl'}+\mathcal{O}(\epsilon^2),
\end{equation}
\begin{equation}
    g=g'+\epsilon\sum_k \frac{\partial S_k(g',L',G')}{\partial G'}\mathrm{e}^{\mathrm{i}kl'}+\mathcal{O}(\epsilon^2),
\end{equation}
\begin{equation}\label{eq:G}
    G=G'+\epsilon\sum_k \frac{S_k(g',L',G')}{\partial g'}\mathrm{e}^{\mathrm{i}kl'}+\mathcal{O}(\epsilon^2).
\end{equation}

The generating function \eqref{eq:generator} is designed such that the new Hamiltonian no longer contains oscillatory terms of order $\epsilon$:
\begin{align}\label{eq:H'}
    &\mathcal{H}'=\nonumber\\&\mathcal{H}_0(L')+\epsilon \overline{\mathcal{H}}_1(L',g',G')+\epsilon^2\left(\overline{\mathcal{H}}_2(L',g',G')+\widetilde{\mathcal{H}}_2(l',L',g',G')\right)+\mathcal{O}(\epsilon^3).
\end{align}
Restricting ourselves to $\mathcal{O}(\epsilon)$ precision, we observe that (i) $\dot{L'}=\partial(\mathcal{H}_0+\epsilon \overline{\mathcal{H}}_1)/\partial l'=0$ and (ii) $L-L'\propto \epsilon \mathrm{e}^{\mathrm{i}kl}$. Given that $L\propto \sqrt{a}$, it follows from (i) and (ii) that the semi-major axis oscillates on orbital timescale and with small amplitude around a constant value. This is the origin of the standard lore that the semi-major axis does not evolve secularly in the absence of resonances.

Von Zeipel's procedure can be continued as above up to arbitrarily high accuracy. In a second-order theory, we eliminate $\epsilon^2\widetilde{\mathcal{H}}_2$ as well. To do so, we apply a generating function very similar to Eq.~\eqref{eq:generator}:
\begin{equation}\label{eq:S'}
    S'=\mathrm{id.}+\epsilon^2\sum_m S_m(g',L'',G'')\mathrm{e}^{\mathrm{i}ml'}
\end{equation}
The resulting expressions are completely analogous to Eqs.~\eqref{eq:l}-\eqref{eq:G}. Expressing the primed variables with double-primed ones in Eq.~\eqref{eq:L} and omitting terms smaller than $\mathcal{O}(\epsilon^3)$ yields 
\begin{align}\label{eq:fullL}
    L=L''+\epsilon^2\sum_m \mathrm{i}mS_m(g'',L'',G'')\mathrm{e}^{\mathrm{i}ml''}+\epsilon\sum_k \mathrm{i}kS_k(g'',L'',G'')\mathrm{e}^{\mathrm{i}kl''}+\nonumber\\\epsilon^3\sum_{k,m}\mathrm{i}k \left( 
    \frac{\partial S_k}{\partial g''}\frac{\partial S_m}{\partial G''}+\frac{\partial S_k}{\partial L''}\mathrm{i}mS_m+\frac{\partial S_k}{\partial G''}\frac{\partial S_m}{\partial g''} \right)\mathrm{e}^{\mathrm{i}(k+m)l''}+\mathcal{O}(\epsilon^4),
\end{align}
while the new Hamiltonian is 
\begin{equation}
    \mathcal{H}''=\mathcal{H}_0(L'')+\epsilon \overline{\mathcal{H}}_1(L'',g'',G'')+\epsilon^2 \overline{\mathcal{H}}_2(L'',g'',G'')+\mathcal{O}(\epsilon^3).
\end{equation}
Note the slight abuse of notation: $S_k(x'')$ or $\partial S_k (x'')$ means that we evaluate the function or its derivative at $x''$, where $x$ is one of the canonical variables. Now we make the following observations: (i) $L''$ is constant while $g''$ and $G''$ evolves secularly; 
(ii) $L-L''$ may contain a term which does not oscillate on orbital timescale, provided $k=-m.$ To put it another way, $a$ may evolve on longer-than-orbital times, underlining the discovery made in Ref.~\cite{conway2024a}.

\section{Toy model}\label{sec:toymodel}
Given the lengthy formulae obtained even in the simple case of a Newtonian hierarchical triple, hereafter we use a simple toy model\footnote{Note that this Hamiltonian simply serves the purpose of building the intuition of the abstract theory above; it does not have a direct physical meaning, although structurally similar terms show up in real Hamiltonians.} to demonstrate the ideas above.

Let us take the following Hamiltonian:
\begin{equation}
    \mathcal{H}=-\frac{1}{2L^2}+\frac{\epsilon}{\mathfrak{D}^3}G \sin l +\frac{\epsilon}{\mathfrak{D}^2}  \sin g,
\end{equation}
where $\mathfrak{D}$ is a constant with the same dimension and similar magnitude as $L$ and $G$. Equation \eqref{eq:generator} now has the form
\begin{equation}
    S=lL' + gG' +\frac{\epsilon L'^3G}{\mathfrak{D}^3}  \cos l,
\end{equation}
which provides the following dictionary between
the old and new variables:
\begin{equation}
    l'=l+\frac{3\epsilon L'^2G'}{\mathfrak{D}^3}  \cos l,
\end{equation}
\begin{equation}
    L=L'-\frac{\epsilon L'^3G'}{\mathfrak{D}^3} \sin l,
\end{equation}
\begin{equation}
    g'=g+\frac{\epsilon L'^3}{\mathfrak{D}^3} \cos l,
\end{equation}
\begin{equation}
    G=G'.
\end{equation}
We express the old variables with the new ones (see Eqs.~\ref{eq:l}-\ref{eq:G}; note that we omit the $\mathcal{O}()$ notations hereafter, so the ``$=$" sign is meant only approximately):
\begin{equation}
    l=l'-\frac{3\epsilon L'^2G'}{\mathfrak{D}^3}\cos l' - \frac{9\epsilon^2 L'^4G'^2}{\mathfrak{D}^6}\sin l' \cos l',
\end{equation}
\begin{equation}\label{eq:originalL}
    L=L'-\frac{\epsilon L'^3G'}{\mathfrak{D}^3} \sin l'+\frac{3\epsilon^2L'^5G'^2}{\mathfrak{D}^6}\cos^2 l' + \frac{27}{2}\frac{\epsilon^3 L'^7G'^3}{\mathfrak{D}^9}\sin l' \cos^2 l' ,
\end{equation}
\begin{equation}
    g=g'-\frac{\epsilon L'^3}{\mathfrak{D}^3} \cos l' - \frac{3\epsilon^2L'^5G'}{\mathfrak{D}^6} \sin l' \cos l',
\end{equation}
\begin{equation}
    G=G'.
\end{equation}
Given that $\partial S /\partial t=0$, the new Hamiltonian is the same as the old one. Expressed with the new variables, it reads (see Eq.~\ref{eq:H'})
\begin{align}\label{eq:transformedham}
    \mathcal{H}'=-\frac{1}{2L'^2}&+\frac{\epsilon }{\mathfrak{D}^2} \sin g' - \frac{\epsilon ^2 L'^3}{\mathfrak{D}^5} \cos g' \cos l' - \frac{3}{2}\frac{\epsilon^2 L'^2G'^2}{\mathfrak{D}^6} \sin^2 l' +\nonumber \\ &+\frac{9\epsilon^3L'^4G'^3}{\mathfrak{D}^9} \sin l'\cos^2l'-\frac{\epsilon^3L'^4G'^3}{\mathfrak{D}^9} \sin^3 l'-\nonumber \\&-\frac{3\epsilon^3L'^5G'}{\mathfrak{D}^8}\cos g' \sin l' \cos l' - \frac{1}{2}\frac{\epsilon^3L'^6}{\mathfrak{D}^8} \sin g' \cos ^2 l'.
\end{align}

As required in a first-order theory, it does not depend on the fast angle $l'$ at order $\epsilon$. Now we perform the second von Zeipel transformation in order to eliminate the oscillatory $\mathcal{O}(\epsilon^2)$ terms in $\mathcal{H}'$. The generating function \eqref{eq:S'} is
\begin{equation}
    S'=\mathrm{id.}+...+\frac{\epsilon^2L''^6}{\mathfrak{D}^5}\cos g' \sin l' +...,
\end{equation}
where, to avoid the clutter in the expressions, we highlighted only one extra term explicitly, which is to eliminate the third term on the r.h.s. of Eq.~\eqref{eq:transformedham}. $S'$ is defined such that the new Hamiltonian is 
\begin{equation}
    \mathcal{H}''=-\frac{1}{2L''^2} + \frac{\epsilon}{\mathfrak{D}^2} \sin g'' - \frac{3}{4}\frac{\epsilon^2L''^2G''^2}{\mathfrak{D}^6} -\frac{1}{4} \frac{\epsilon^3L''^6}{\mathfrak{D}^8}\sin g''.
\end{equation}
Using the generating function $S'$ and the simplified notations, we get
\begin{equation}
    G'=G''+...-\frac{\epsilon^2L''^6}{\mathfrak{D}^5} \sin g' \sin l' + ...
\end{equation}
Substituting it back into the second term in Eq.~\eqref{eq:originalL} yields
\begin{align}\label{eq:newL}
    L&=L''+...-\epsilon \frac{L''^3}{\mathfrak{D}^3}\left(G''+...-\frac{\epsilon^2L''^6}{\mathfrak{D}^5} \sin g'' \sin l'' + ...\right)\sin l''+...\nonumber \\&=L''+...+\frac{\epsilon^3L''^9}{\mathfrak{D}^8}\sin g''+...,
\end{align}
which is analogous to Eq.~\eqref{eq:fullL}. Note that the term highlighted does not depend on the fast angle $l''$. It is purely secular and driven by $\mathcal{H}''$: $L''$ is constant while $g''$ evolves secularly according to
\begin{equation}
    \dot{g''}=\frac{\partial \mathcal{H}''}{\partial G ''}=-\frac{3}{2}\frac{\epsilon^2L''^2G''}{\mathfrak{D}^6}.
\end{equation}
Since $L=\sqrt{a}$ is related to the original (and thus ``true") semi-major axis, we conclude that the semi-major axis evolves secularly.

\section{Discussion and conclusion}\label{sec:discussion}
As first-order quadrupole theory suffices in most of the applications, most studies never apply the canonical transformation generated by \eqref{eq:S'}. Since the secular term in Eq.~\eqref{eq:newL} is the result of substituting the higher-order terms into the lower-order ones, the semi-major axis never evolves secularly in a first-order theory (assuming there are no mean motion resonances). 
We close the discussion with two remarks. First, in the presentation above the semi-major axis' secular evolution is proportional to $\epsilon^3$, while in Ref.~\cite{conway2024a} to $\epsilon^2$. The difference most likely stems from the fact that they apply the $\mathcal{O}(\epsilon)$ generating function twice, (i.e., their $\mathcal{H}$ is doubly averaged), thus $\mathrm{e}^{\mathrm{i}(k+m)l}$-like terms may appear at lower order. Second, such secular evolution in the semi-major axis does not appear if the perturbation (and hence the generating function) is such that the prefactor of $\mathrm{e}^{\mathrm{i}(k+m)l''}$ in Eq.~\eqref{eq:fullL} is zero.  

\section*{Acknowledgments}
The author acknowledges the financial support of the Hungarian National
Research, Development and Innovation Office -- NKFIH Grant OTKA K-147131.

\bibliographystyle{unsrt}
\bibliography{cit}

\end{document}